\documentclass[]{raa}
\usepackage{amsfonts}
\usepackage{mathrsfs}
\usepackage{txfonts}
\usepackage{amssymb}            
\usepackage{graphicx,times}
\usepackage{natbib}

\providecommand{\bm}[1]{\mbox{\boldmath$#1$\unboldmath}}

\begin{document}

   \title{A multiwavelength study of the star forming HII region Sh2-82
}

 \volnopage{ {\bf 20xx} Vol.\ {\bf 9} No. {\bf XX}, 000--000}
   \setcounter{page}{1}

   \author{Naiping Yu\inst{1,2}, Jun-Jie Wang\inst{1,2}
   }

   \institute{National Astronomical Observatories, Chinese Academy of Sciences,
             Beijing 100012, China; {\it yunaiping09@mails.gucas.ac.cn}\\
        \and
             NAOC-TU Joint Center for Astrophysics, Lhasa 850000, China
        \and
\vs \no
   {\small Received [year] [month] [day]; accepted [year] [month] [day]
} }

\abstract{ Based on a multiwavelength study, the interstellar medium
and young stellar objects (YSOs) around the HII region Sh2-82 have
been analyzed. Two molecular clumps were found from the archival
data of the Galactic Ring Survey, and using the Two Micron All-Sky
Survey catalog, we found two corresponding young clusters embedded
in the molecular clumps. The very good relations between CO
emission, infrared shells and YSOs suggest that it is probably a
triggered star formation region from the expansion of Sh2-82. We
further used the data from the Galactic Legacy Infrared Mid-Plane
Survey Extraordinaire from Spitzer to study the YSOs within the two
clumps, confirming star formation in this region. By spectral energy
distribution fits to each YSO candidate with infrared access, we
derived the slope of the initial mass function. Finally, comparing
the HII region's dynamical age and the fragmentation time of the
molecular shell, we discard the ``collect and collapse'' process as
being the triggering mechanism for YSO formation. Sh2-82 can be a
mixture of other processes such as radiative-driven implosion and/or
collisions with pre-existing clumps.  }

   \authorrunning{Naiping Yu, Jun-Jie Wang }            
   \titlerunning{A multiwavelength study of the star formation HII region Sh2-82}  
   \maketitle


%
%
\section{INTRODUCTIONS}           

Given that our own solar system probably formed in a massive cluster
(Hester et al. 2004), knowledge of triggered star formation is
necessary for understanding our own formation. Massive stars can
strongly influence their surrounding environment via stellar winds,
ionizing radiation and the expansion of HII regions. Since Elmegreen
\& Lada (1977) presented a mechanism for the sequential star
formation of OB subgroups in molecular clouds, a number of
mechanisms by which massive stars can affect the subsequent star
formation around HII regions have been proposed. Two of the most
studied mechanisms for triggering star formation in HII regions are
``radiatively driven implosion'' (RDI) (e.g. Lefloch \& Lazareff
1994; Miao et al.2006; Miao et al. 2009) and ``collect and
collapse'' (C\&C) (Elmegreen \& Lada 1977). In the RDI model,
pre-existing molecular clumps in the cloud become surrounded on all
sides by high-pressure ionized gas heated by UV radiation, leading
to the formation of a cometary globule (Bertoldi \& McKee 1990). A
dense core will form in the cometary globule where star formation
will finally take place. In the C\&C model, a slow-moving D-type
ionization front has an associated shock front that precedes the
ionization front. Dense gas may pile up between the two fronts. Over
a long time the compressed shocked layer becomes gravitationally
unstable and fragments into dense clumps.
\begin{figure} [!h]
\onecolumn \centering
\includegraphics[width=6in,height=4in]{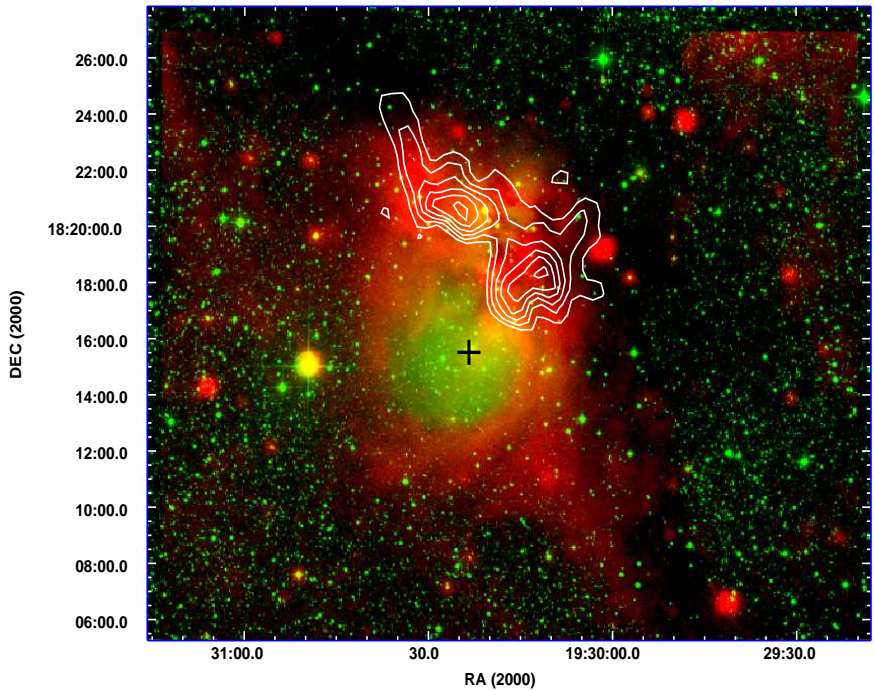}

\caption{$^{13}$CO (1-0) integrated emission between 20.5 km
s$^{-1}$ and 27.5 km s$^{-1}$ superimposed on the DSS (green) and
MSX A band (red) pictures. The contour levels are from 8 K km
s$^{-1}$ to 18 K km s$^{-1}$ by step of 2 K km s$^{-1}$. The black
cross represents the ionizing star HD 231616.}
\end{figure}

  With the aim of increasing the observational evidence of triggered star formation in the
surroundings of HII regions, we present a multiwavelength study of
the molecular environment and YSOs of the HII region Sh2-82. Sh2-82
is ionized by HD 231616, a B0V/III star with a mass of 18
M$_{\odot}$ (Hunter $\&$ Massey 1990) (Fig.1). The distance of
Sh2-82 can be estimated in several ways. Based on UBVRI
photoelectric photometry, Lahulla (1985) derived a mean distance of
1.58 kpc, with the data ranging from 1.40 kpc to 1.80 kpc. From
simulated B and V photometry formed through IIDS spectra of star(s),
Hunter $\&$ Massey (1990) derived the distance of the ionizing star
in Sh2-82 to be 1.7 Kpc. According to the Galactic rotation model of
Brand $\&$ Blitz (1993) (with R$_{\odot}$=8.2 kpc and
$\upsilon$$_{\odot}$=220 km s$^{-1}$ ), we obtain a kinematic
distance of either 2.1 kpc or 8.5 kpc. This ambiguity arises because
we are studying a region in the first Galactic quadrant, where a
given velocity may be associated with two possible distances.
Considering that distances derived by photometry are more accurate
than the other two methods, we use a distance of 1.8 $\pm$ 0.4 kpc
for Sh2-82 in the following discussion.

\section{DATA SETS, REDUCTIONS, AND RESULTS}

\subsection{GRS}
We analyzed the radio emission of $^{13}$CO (1-0) in the region of
Sh2-82 using the Galactic Ring Survey (GRS; Jackson et al. 2006).
The survey used the SEQUOIA multi pixel array on the Five College
Radio Astronomy Observatory's 14 m telescope to cover a longitude
range of $\ell$ = 18$^{\circ}$$\sim$55.7$^{\circ}$ and a latitude
range of $\mid$b$\mid$ $<$ 1$^{\circ}$, fully sampled with a pixel
size of 22 arcsec from 1998 to 2005. The survey's velocity coverage
is -5 to 135 km s$^{-1}$ for Galactic longitudes $\ell$ $\leq$
40$^{\circ}$and -5 to 85 km s$^{-1}$ for Galactic longitudes $\ell$
$>$ 40$^{\circ}$. The spectral resolution for both velocity ranges
is 0.21 km s$^{-1}$. The data cube was reduced with IDL procedures.

  Figure 1 shows a distinctive ring morphology in the MSX A band detection. The 8.3$\mu$m emission
is dominated by intense line emission from polycyclic aromatic
hydrocarbons (PAHs), arising at the interface between ionized gas in
the HII region and the surrounding molecular material. We inspected
the molecular gas around Sh2-82 from the GRS data in the whole
velocity range and found an interesting feature around $\upsilon$
$\sim$ 24 km s$^{-1}$ (figure 2). Two molecular clumps are evident
on the northern photo dominated region (PDR) bordering Sh2-82. The
very good correspondence between the HII region border traced by the
IR emission and the molecular gas, strongly suggests that the
observed molecular gas is associated with Sh2-82. We used the
central velocity of the molecular gas to infer its kinematic
distance.

\begin{figure} [!h]
\onecolumn \centering
\includegraphics[width=6.5in,height=3in]{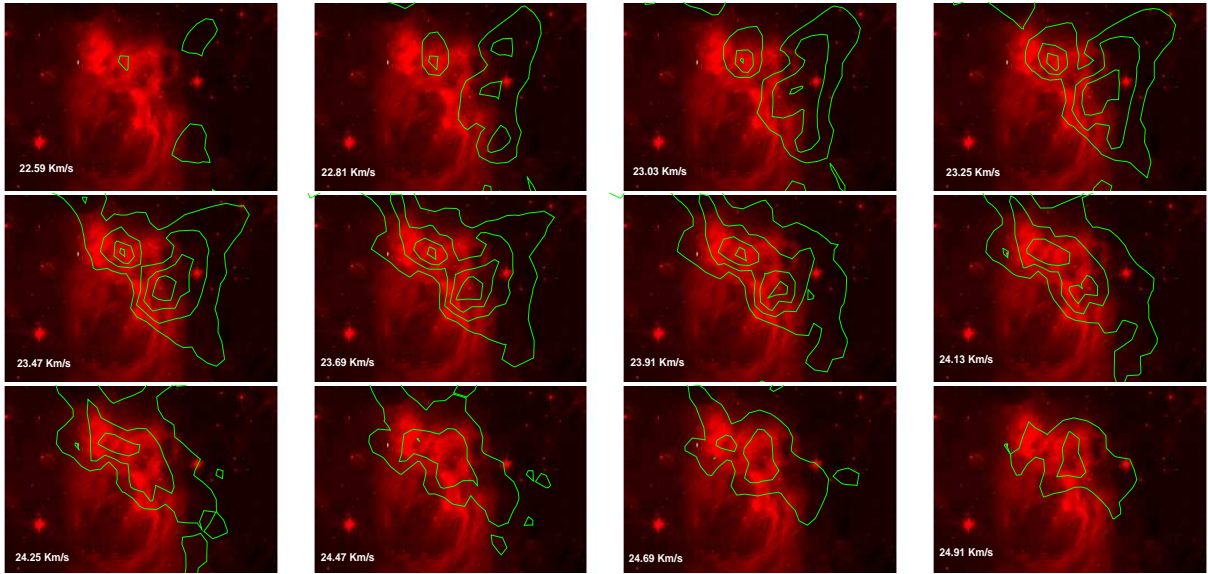}

\caption{Channel maps of the 13CO (1-0) emission every 0.22km
s$^{-1}$ superimposed on the Spiter 8.0$\mu$m image. The green
contour levels are 2, 4, 6 and 8 K km s$^{-1}$.}
\end{figure}

Using the $^{13}$CO (1-0) line and under the assumption of local
thermodynamic equilibrium (LTE), the H$_{2}$ column density and the
masses of the two clumps can be estimated. We use
\begin{equation}
N(^{13}CO)=2.4\times10^{14}\frac{T_{ex}\int \tau_{13}
d\upsilon}{1-exp(-5.29/T_{ex})}
\end{equation}
to obtain the column density, where $\tau _{13}$ is the optical
depth. We assume that the $^{13}$CO emission is optically thin, the
excitation temperature T$_{ex}$=20K and the solar abundance ratios
[H$_{2}$]/ [$^{12}$CO] =10$^{4}$ and [CO]/ [$^{13}$CO] =89(Wilson \&
Rood 1994). The molecular masses were obtained from
\begin{equation}
M[M_{\odot}]=4.2\times10^{-20} N(H_{2}) D^{2} A
\end{equation}
where D is the distance in pc and A is the solid angle in
steradians. The masses we derived were 2400M$_{\odot}$ and
8600M$_{\odot}$, respectively, for the two clumps.

\begin{figure} [htbp]
\onecolumn \centering
\includegraphics[width=4.5in,height=3.5in]{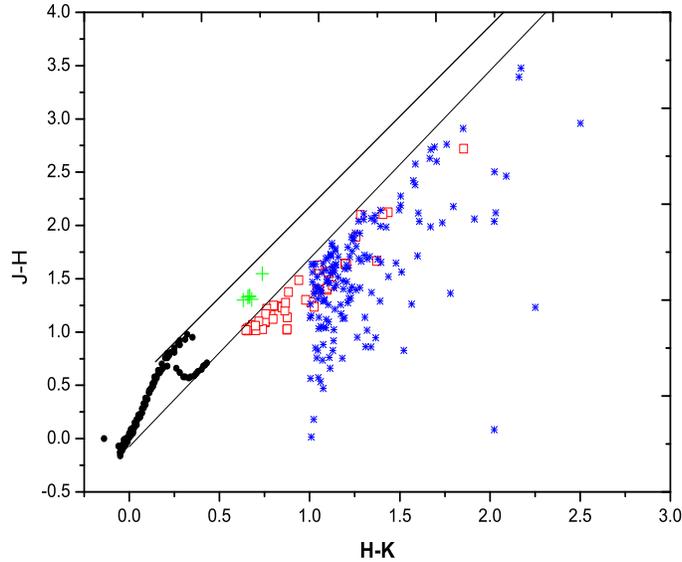}
\caption{Color-Color diagram of the 2MASS YSO candidates projected
on to Sh2-82. The P$_{1}$, P$_{1+}$ and P$_{2}$ sources are
indicated by red boxes, green pluses and blue asterisks,
respectively. }
\end{figure}

\begin{figure}[h]
\onecolumn \centering
\begin{center}
\includegraphics[width=2.3in,height=2.3in]{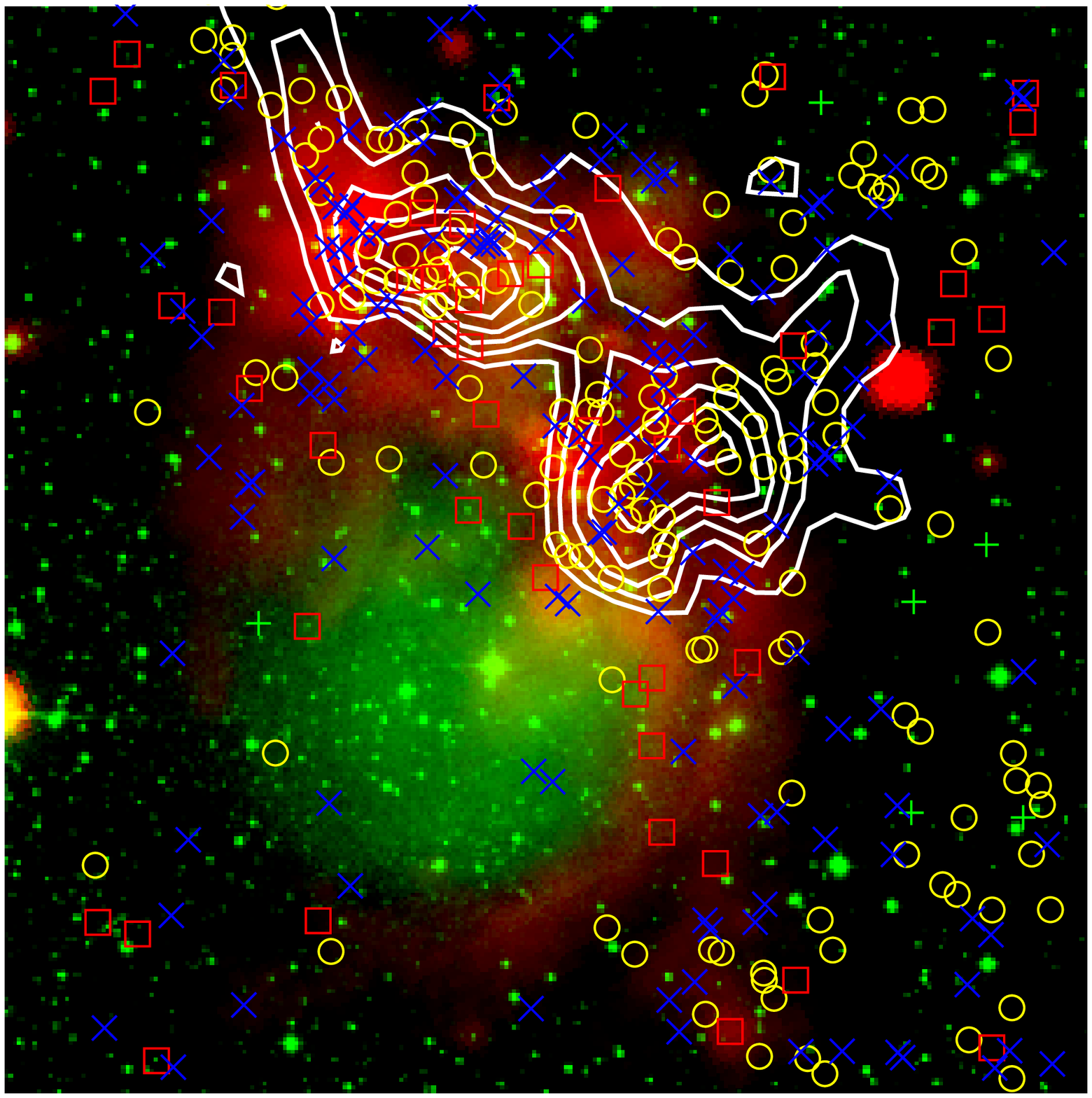}
\end{center}
\begin{center}
\includegraphics[width=2.5in,height=3.2in,angle=270]{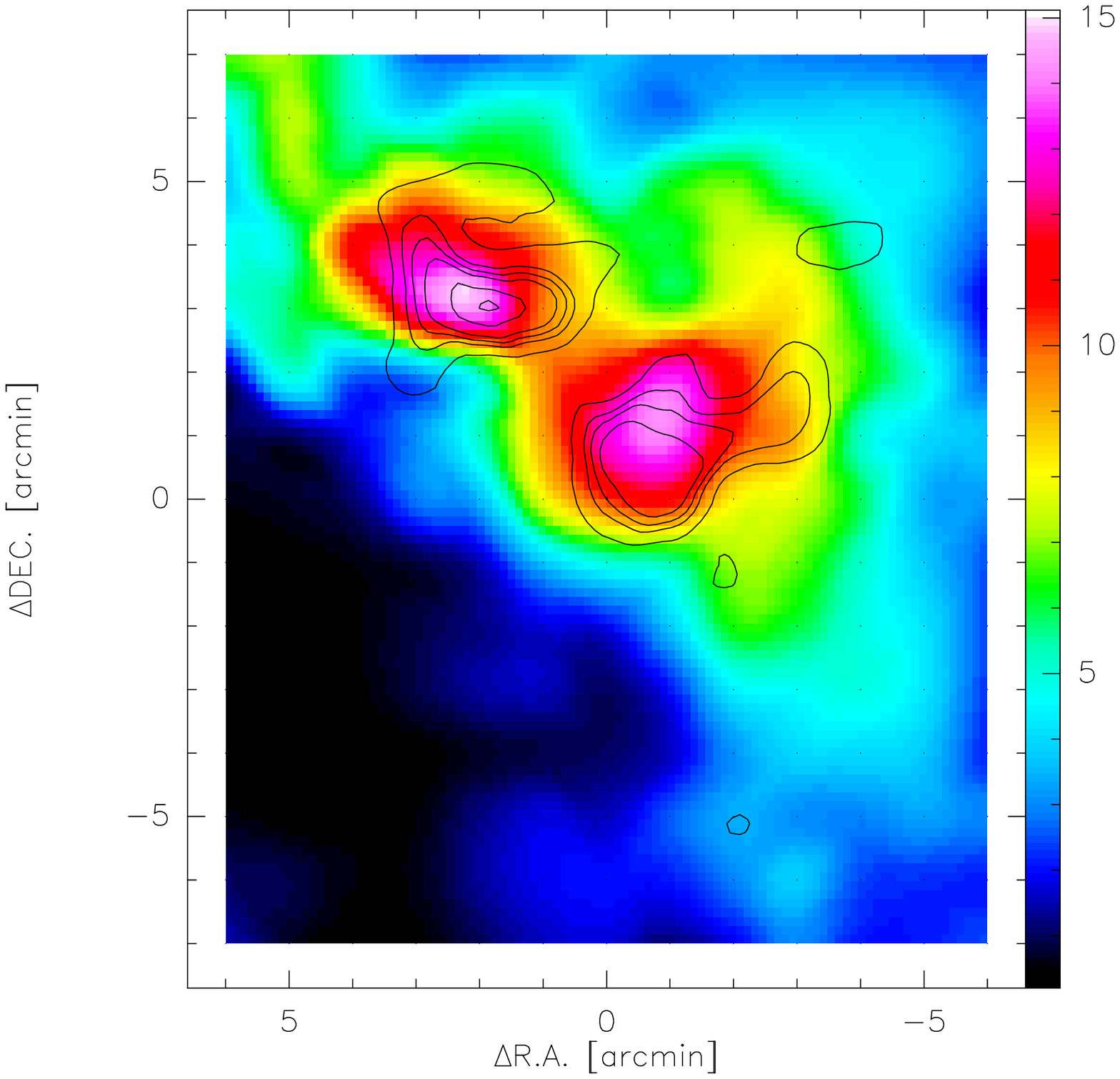}
\end{center}
\caption{ The two young clusters found through the 2MASS point
catalog. Up: P$_{1}$, P$_{1+}$, P$_{2}$ and P$_{3}$ sources
indicated by red boxes, green pluses, blue crosses and yellow
circles, respectively. The contour levels are the same as figure 1
for the $^{13}$CO(J=1-0) emission. Down: contours corresponding to
stellar density ranges of 5, 7, 8, 10 and 12arcmin$^{-2}$
superimposed onto the emission intensity of $^{13}$CO (1-0).}
\end{figure}
\subsection{2MASS point sources}
We searched the environment of Sh2-82 for IR star clusters to locate
sites of recent star formation. Based on a sample of YSOs having low
and intermediate masses, Kerton et al. (2008) inferred color
criteria to analyze the presence of star formation activity in the
surroundings of the HII region KR 140. Kerton et al. (2008) divided
their YSO candidates into four groups (P$_{1}$, P$_{1+}$, P$_{2}$,
P$_{3}$) according to different photometric qualities and the
adopted color criteria. P$_{1}$ sources have valid photometry in all
three bands (i.e. ph qual values = A, B, C or D). The colour
criteria are (J-H)$>$1 and (J-H)-1.7(H-K)+0.075$<$0. This approach
selects stars lying below the reddening vector associated with an
O6V star. P$_{1+}$ sources also have valid photometry in all three
bands. The color criteria are 1.3$<$(J-H)$<$1.6,
(J-H)-1.7(H-K)+0.075$>$0, (J-H)-1.7(H-K)-0.3805$<$0 and K$>$14.5.
This method selects YSOs lying in the overlapping region of T Tauri
and main sequence stars. The P$_{2}$ group has not been detected in
the J band. Thus the actual position of P$_{2}$ sources in the (J-H)
axes should be towards higher values. The P$_{2}$ color criteria are
(J-H)-1.7(H-K)+0.075$<$0 and (H-K)$>$1.0. Sources belonging to the
P$_{3}$ group have J and H magnitudes that are lower limits, so
their color (J-H) cannot be considered. The P$_{3}$ color criterion
is (H-K)$>$1.0. Following such criteria, we searched for tracers of
stellar formation activity in the 2MASS catalog, finding two YSO
clusters associated with each molecular clump (Fig 4).

Figure 3 shows the color-color diagram for the YSO candidates. The
two parallel lines are reddening vectors, assuming the interstellar
reddening law of Rieke \& Lebofsky (1985) (A$_{J}$ / A$_{V}$ =
0.282; A$_{H}$ / A$_{V}$ = 0. 175; A$_{K}$ / A$_{V}$ = 0.112).
Figure 4 shows the two young clusters embedded in the two molecular
clumps.

\subsection{Spitzer Data}
The Galactic Legacy Infrared Mid-Plane Survey Extraordinaire
(GLIMPSE I; Benjamin et al. 2003) covered the Galactic plane
(10$^{\circ}$$<$$\mid$$\ell$$\mid$$<$65$^{\circ}$,
$\mid$b$\mid$$<$1$^{\circ}$) with the four mid-IR bands (3.6, 4.5,
5.8, 8.0$\mu$m) of the Infrared Array Camera (IRAC; Fazio et al.
2004) on the Spitzer Space Telescope. IRAC has a resolution of
1.5${''}$-1.9${''}$ (3.6-8.0$\mu$m). In this region we used the
highly reliable GLIMPSE I Catalogue with the sources detected in
four bands (The highly reliable GLIMPSE I, II and 3D catalogs
consists of point sources that are detected at least twice in one
band with S/N $>$ 5, and at least once in an adjacent band). The
GLIMPSE I Catalog also tabulates JHK$_{S}$ flux densities from the
2MASS point source catalogs (Skrutskie et al. 2006) for all GLIMPSE
sources with 2MASS identifications.

MIPSGAL (Carey et al. 2005) is a legacy program covering the inner
Galactic plane 10$^{\circ}$$<$$\mid$$\ell$$\mid$$<$65$^{\circ}$,
$\mid$b$\mid$$<$1$^{\circ}$ at 24 and 70$\mu$m with the Multiband
Imaging Photometer on Spitzer (MIPS; Rieke et al. 2004). It has a
resolution of 6${''}$ at 24$\mu$m and 18${''}$ at 70$\mu$m. MIPSGAL
point-source catalogs were not yet available at the time of this
study. For the 24$\mu$m band, we used the APEX 1-Frame routines and
the point response function (PRF) provided on the Spitzer Science
Center's web site to perform point-source PRF-fitting photometry. If
an archive source was located within 2.4${''}$ of a 24$\mu$m source,
it was considered to have a 24$\mu$m detection. For sources that
APEX failed to detect automatically, we used the user list option in
APEX to supply the coordinates for these sources to successfully
derive a PSF fit.

With highly reliable infrared detections, we can distinguish YSOs
from normal stars, i.e. main-sequence, giant and supergiant stars,
from their excess IR emission, as they are enshrouded in dust that
absorbs stellar UV and optical radiation. The regions in the left
image of figure 5 indicate stellar evolutionary stages based on the
criteria described by Allen et al. (2004): Class I sources are
protostars with interstellar envelopes and Class II sources are
disk-dominated objects. However, background galaxies and evolved
stars such as asymptotic giant branch (AGB) stars can also be red
sources. In order to make a more pure YSO candidate sample, we
removed star-formation (PAH) galaxies and weak-line AGNs via a
series of cuts in the four-band IRAC color-color diagrams using the
procedure developed by Gutermuth et al. (2008). The color criteria
for both AGB stars and extragalactic contaminants from Harvey et al.
(2006) were also used to filter our sample.

Figure 5  right image) shows the final distributions of both the
Class I (magenta pluses) and Class II (cyan pluses) point sources
around Sh2-82. It can be noted that most YSOs prefer to be located
on the northwest corner of Sh2-82, and nearly all Class I sources
are projected onto the molecular clumps. Since molecular clouds are
birth places of stars, we expect some of these sources to be YSOs
whose formation could have been triggered by the expansion of
Sh2-82. We also perform a fitting of these YSO candidates in the
IRAC and 2MASS bands to derive their spectral energy distribution
(SED). We limit our study to the sources superimposed on the two
molecular clumps. Briefly, the SED-fitting tool works as a
regression method to find the SEDs within a specified $\chi$$^{2}$
from a large grid of models after fitting the input data points. The
grid of models contains stellar masses, disk masses, mass accretion
rates, and line-of-sight (LOS) inclinations. The grid of the YSO
models was computed by Robitaille et al. (2006) using the 20 000
two-dimensional radiation transfer models from Whitney et al.
(2003a, 2003b, 2004). Each YSO model has SEDs for 10 viewing angles
(inclinations), so the total YSO grid consists of 200 000 SEDs.
Observations from the J-band to 24$\mu$m are fitted using a
$\chi$$^{2}$-minimization technique. We consider sources with
$\chi$$^{2}$/N$_{data}$ $<$ 4 to be well fitted. Our results are
listed in table 1.

Figure 6 shows two examples of the SED fitting results. From these
fits, we can see that there is a large infrared excess, which is
inferred to be a planetary disk around the central stars (Wang \&
Hu, 1994). None of the selected YSO candidates were better fitted by
a pure stellar photosphere model, confirming our YSO color selection
criteria for this region.

\begin{figure}
\onecolumn \centering
\includegraphics[height=2.4in,width=3.2in]{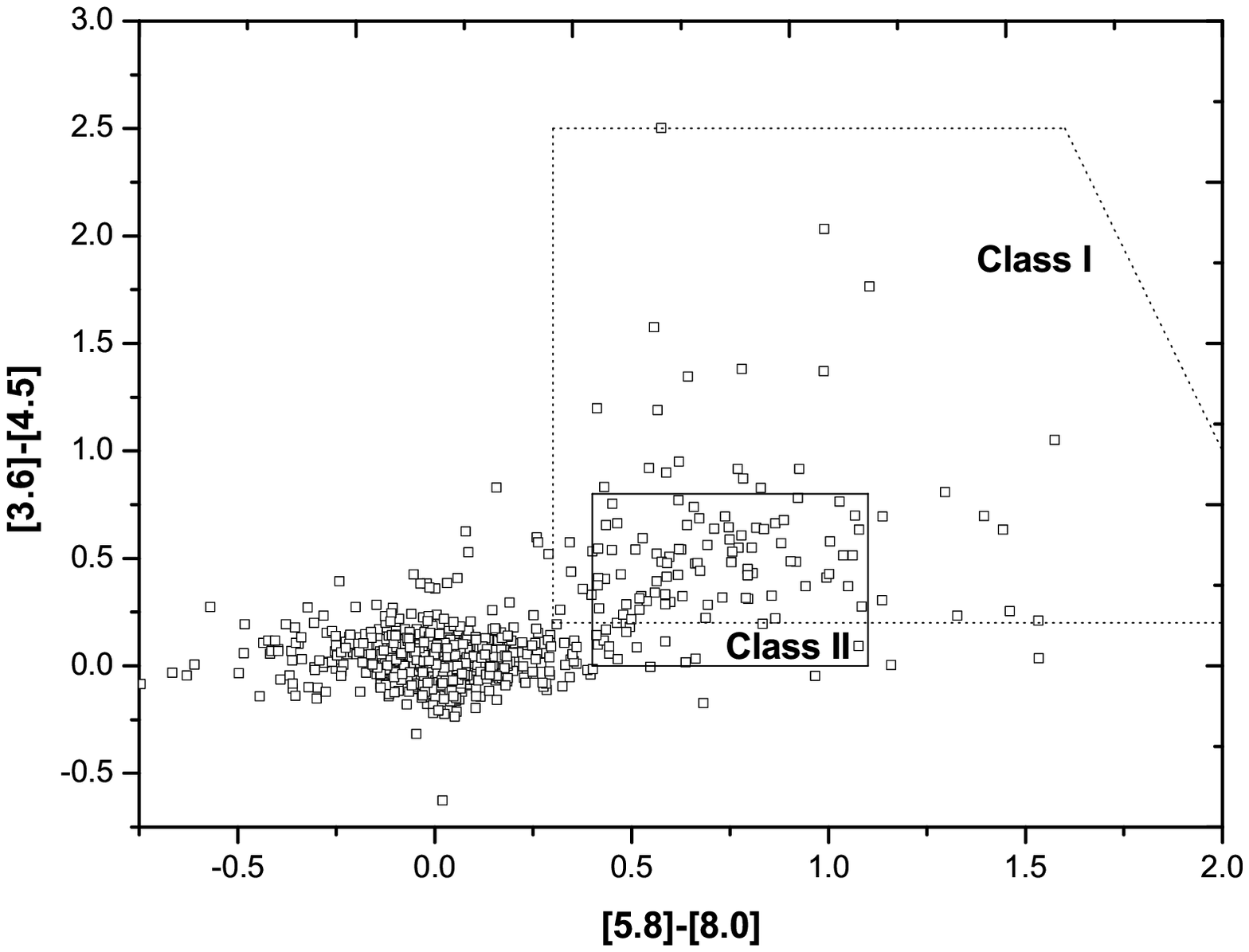}
\includegraphics[width=2.3in,height=2.2in]{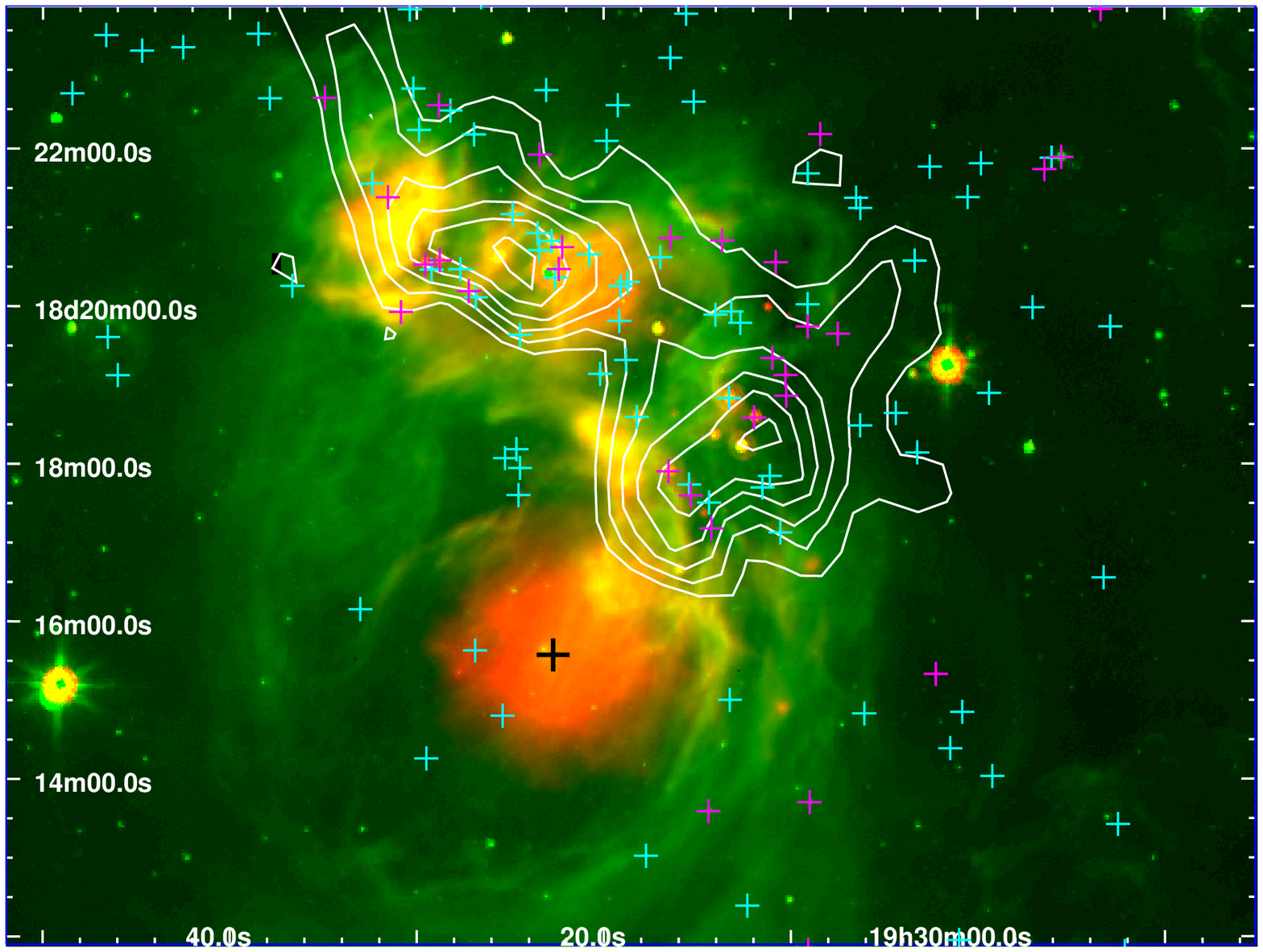}
\caption{ Left: The distribution of YSOs within the color selection
criteria of Allen et al(2004). Right: a two color image of Sh2-82:
8$\mu$m emission (in green) and 24$\mu$m emission (in red). The
magenta pluses are Class I and the cyan pluses are Class II YSOs.
The black plus is the ionizing star of this HII region.}
\end{figure}

\begin{figure}
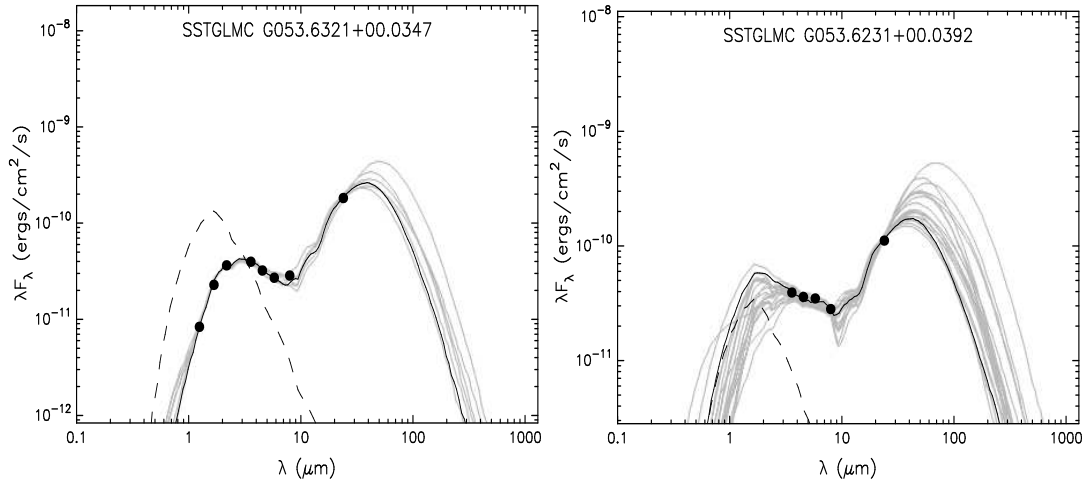

\onecolumn \centering
\includegraphics[width=2.80in,height=2.50in]{example1figure6.eps}
\includegraphics[width=2.80in,height=2.50in]{example2figure6.eps}
\caption{ Two examples of the SED fitting model. The dashed line
represents the stellar photosphere model. The black line represents
the best-fitting SED, and the gray lines represent all the other
acceptable YSO fits.}
\end{figure}

\begin{table*}
\begin{minipage}{15cm}
 \caption{\label{tab:test} Parameters derived from the SED fitting of sources projected onto the molecular clumps. }

\begin{tabular}{lclclclclclclcl}
 \hline
    Source name  &  R.A.(J2000) &  DEC(J2000)  &  M(M$_{\odot}$) &  $\chi$$^{2}$ (total) & M($_{disk}$)(M$_{\odot}$) & $\dot{M}$$_{env}$(M$_{\odot}$/yr) \\
\hline
 SSTGLMC G053.6168+00.0448 & 292.5866  &  18.34431  &  3.72  &  5.65  &  2.15 $\times$ 10$^{-4}$  &  0.00     \\
 SSTGLMC G053.6209+00.0406 &  292.5926  &  18.34584  &  1.82  &  5.04  &  6.52 $\times$ 10$^{-3}$  &  9.96 $\times$ 10$^{-6}$    \\
 SSTGLMC G053.6172+00.0377  & 292.5934   & 18.34121   & 7.49   & 18 & 4.52 $\times$ 10$^{-2}$  &  5.56 $\times$ 10$^{-5}$         \\
 SSTGLMC G053.6160+00.0361  & 292.5942   & 18.33941  & 7.29 & 17.65  & 1.24 $\times$ 10$^{-2}$  & 3.04 $\times$ 10$^{-5}$ \\
 SSTGLMC G053.6231+00.0392  & 292.595  & 18.34714 &   2.21 &   2.6 & 9.16 $\times$ 10$^{-2}$   & 4.49 $\times$ 10$^{-6}$ \\
\\
 SSTGLMC G053.6180+00.0352  & 292.596  & 18.34073  &  4.89  &  16.52  & 7.06 $\times$ 10$^{-2}$  & 1.32 $\times$ 10$^{-4}$ \\
 SSTGLMC G053.6404+00.0458  & 292.5976  &  18.36544 &   1.22 &   0.5 & 9.16 $\times$ 10$^{-4}$ & 9.01 $\times$ 10$^{-6}$ \\
 SSTGLMC G053.6227+00.0360  & 292.5977   & 18.34522 &   4.56 &   3.35  &  7.01 $\times$ 10$^{-4}$ & 2.21 $\times$ 10$^{-5}$ \\
 SSTGLMC G053.6260+00.0374  & 292.5981  &  18.34881  &  1.22  &  8.03 &   1.25 $\times$ 10$^{-2}$ & 1.1 $\times$ 10$^{-5}$ \\
 SSTGLMC G053.6089+00.0238 &  292.602 & 18.32729 &   3.36 &   5.58 &   1.28 $\times$ 10$^{-3}$ & 1.06 $\times$ 10$^{-5}$ \\
\\
 SSTGLMC G053.6321+00.0347  & 292.6036 &   18.35283 &   3.37  &  11.8 &   2.38 $\times$ 10$^{-3}$ & 8.12 $\times$ 10$^{-6}$ \\
 SSTGLMC G053.6508+00.0357  & 292.6122   & 18.36969  &  1.41 &   9.20   & 6.95 $\times$ 10$^{-4}$ & 1.86 $\times$ 10$^{-5}$ \\
 SSTGLMC G053.6222+00.0188  & 292.6134  &  18.33653  &  3.84  &  6.19   & 5.68 $\times$ 10$^{-4}$& 9.85 $\times$ 10$^{-6}$ \\
 SSTGLMC G053.6272+00.0194  & 292.6153   & 18.34119  &  4.63 &   13.93  & 1.10 $\times$ 10$^{-3}$ & 1.43 $\times$ 10$^{-7}$ \\
 SSTGLMC G053.6576+00.0337 &  292.6175  &  18.37475   & 3.15  &  3.61  &  2.76 $\times$ 10$^{-6}$ & 0.00 \\
\\
 SSTGLMC G053.6309+00.0165  & 292.6199   & 18.34309 &   5.43  &  9.91  &  4.47 $\times$ 10$^{-4}$ & 4.15 $\times$ 10$^{-4}$ \\
 SSTGLMC G053.6300+00.0140  & 292.6217 &   18.34109   & 2.92   & 6.00  &  6.80 $\times$ 10$^{-2}$ & 1.24 $\times$ 10$^{-6}$ \\
 SSTGLMC G053.6316+00.0134 &  292.6231   & 18.34215   & 0.17  &  3.56   & 1.81 $\times$ 10$^{-2}$ & 4.73 $\times$ 10$^{-6}$ \\
 SSTGLMC G053.6572+00.0259  & 292.6245  &  18.37064   & 1.77   & 3.12  &  6.33 $\times$ 10$^{-3}$ & 1.46 $\times$ 10$^{-6}$ \\
 SSTGLMC G053.6654+00.0291  & 292.6258  &  18.37940  & 1.32  &  5.81  &  3.84 $\times$ 10$^{-4}$ & 7.25 $\times$ 10$^{-8}$  \\
\\
 SSTGLMC G053.6253+00.0040  & 292.6285  &  18.33213 &   0.34 &   0.6 & 1.72 $\times$ 10$^{-4}$ & 5.60 $\times$ 10$^{-6}$ \\
SSTGLMC G053.6479+00.0133  & 292.6314  &  18.35639 &   5.3 & 6.99  &  3.08 $\times$ 10$^{-2}$ & 3.46 $\times$ 10$^{-4}$ \\
 SSTGLMC G053.6520+00.0118 &  292.6349  &  18.35930 & 1.87 & 1.49   & 1.69 $\times$ 10$^{-2}$ & 2.76 $\times$ 10$^{-5}$ \\
 SSTGLMC G053.6411-00.0133  & 292.6527 &   18.33765  &  1.83 & 8.47  &  1.38 $\times$ 10$^{-3}$ & 5.63 $\times$ 10$^{-5}$ \\
 SSTGLMC G053.5813+00.0781   & 292.5379  &  18.32906  &  3.43  &  0.57  &  1.18 $\times$ 10$^{-3}$ & 0.00 \\
\\
 SSTGLMC G053.5854+00.0803 &   292.5379  &  18.33374  & 0.27  & 1.12  & 2.89 $\times$ 10$^{-3}$& 5.78 $\times$ 10$^{-6}$ \\
 SSTGLMC G053.5706+00.0670  & 292.5427  &  18.31441  &  1.2 & 2.63 & 9.84 $\times$ 10$^{-2}$ & 1.31 $\times$ 10$^{-5}$ \\
 SSTGLMC G053.5745+00.0690  &  292.5428  &  18.31878 &  0.52  & 3.13 & 2.70 $\times$ 10$^{-3}$ & 3.86 $\times$ 10$^{-6}$ \\
 SSTGLMC G053.5790+00.0683  &  292.5457  &  18.32236  &  0.13  & 0.48  & 1.50 $\times$ 10$^{-4}$  & 8.59 $\times$ 10$^{-7}$ \\
 SSTGLMC G053.5573+00.0559  &  292.5463  &  18.29744  &  1.31  & 5.07  & 4.79 $\times$ 10$^{-2}$ & 1.79 $\times$ 10$^{-6}$ \\
\\
 SSTGLMC G053.5560+00.0533   & 292.548  & 18.29500 & 2.09 &   9.22   & 2.261 $\times$ 10$^{-3}$ &  5.19 $\times$ 10$^{-6}$ \\
 SSTGLMC G053.5888+00.0659   & 292.5529  &  18.32982  &  2.01  &  0.84   & 2.88 $\times$ 10$^{-2}$ &  7.35 $\times$ 10$^{-4}$ \\
 SSTGLMC G053.5918+00.0654  &  292.5549  &  18.33224  &  2.16  & 2.85 &   5.10 $\times$ 10$^{-4}$ & 2.73 $\times$ 10$^{-5}$\\
   SSTGLMC G053.5760+00.0562  &  292.5554  & 18.31393  &  6.5  & 28 & 2.56 $\times$ 10$^{-6}$ & 0.00 \\
   SSTGLMC G053.5928+00.0622  &  292.5584  &  18.33156  &  2.49  &  2.43   & 1.33 $\times$ 10$^{-2}$ &  8.37 $\times$ 10$^{-6}$ \\
\\
   SSTGLMC G053.5535+00.0397  & 292.5593  & 18.28632  &  2.95  &  1.51  &  4.21 $\times$ 10$^{-3}$ &  0.00 \\
   SSTGLMC G053.5586+00.0419  &  292.5599   & 18.29182  &  3.21  &  2.00  &  1.45 $\times$ 10$^{-1}$ & 1.13 $\times$ 10$^{-4}$ \\
   SSTGLMC G053.5640+00.0400   & 292.5643   & 18.29563  &  5.63  &  0.23  &  3.60 $\times$ 10$^{-2}$ &  7.73 $\times$ 10$^{-5}$ \\
   SSTGLMC G053.5685+00.0375  &  292.5689  &  18.29842  &  5.65  &  2.23   & 2.62 $\times$ 10$^{-3}$ &  1.25 $\times$ 10$^{-5}$ \\
   SSTGLMC G053.6090+00.0578  &  292.5707  &  18.34370 & 0.28   & 1.32   & 6.74 $\times$ 10$^{-4}$ &  5.05 $\times$ 10$^{-6}$ \\
\\
   SSTGLMC G053.5818+00.0372  &  292.5759   & 18.30991  &  3.34  &  8.59  &  2.34 $\times$ 10$^{-3}$ & 2.55 $\times$ 10$^{-5}$ \\
   SSTGLMC G053.6078+00.0492  &  292.578 & 18.33848  &  1.03  &  5.36   & 2.11 $\times$ 10$^{-4}$ & 6.46 $\times$ 10$^{-6}$ \\
   SSTGLMC G053.5935+00.0409  &  292.5784  &  18.32195  &  2.69  &  4.17  &  1.73 $\times$ 10$^{-2}$ &  4.65 $\times$ 10$^{-5}$ \\
   SSTGLMC G053.6077+00.0474   & 292.5796   & 18.33759  &  0.42 &   2.27  &  2.05 $\times$ 10$^{-2}$ &  1.93 $\times$ 10$^{-4}$ \\
   SSTGLMC G053.6014+00.0437  &  292.5799  &  18.33025  &  4.96  &  9.44   & 7.44 $\times$ 10$^{-5}$ &  5.05 $\times$ 10$^{-6}$ \\
\\
   SSTGLMC G053.5935+00.0347   & 292.5841  &  18.31903   & 4.02  &  16.71  & 6.17 $\times$ 10$^{-3}$ & 3.40 $\times$ 10$^{-8}$ \\
\hline
 \end{tabular}
 \label{tb:rotn}
 \end{minipage}
\end{table*}

If we assume the following form of a mass function:
\begin{equation}
\frac{dN}{dM}=A (M / M_{\odot})^{-\Gamma}
\end{equation}
The distribution of masses of YSOs in Sh2-82 can be estimated. By
studying M17, Povich et al. (2009) found that the GLIMPSE Point
Source Archive recovers essentially all YSOs above $\sim$
3M$_{\odot}$ toward those regions. As is also apparent in figure 7,
the deviation from a single power law becomes progressively worse in
lower mass bin. Using the highest four mass bins (from
3.16M$_{\odot}$ to 8M$_{\odot}$) yields a slope of $\Gamma$=2.21
$\pm$ 0.45, which is approximately consistent with the classical
value of 2.35 derived by Salpeter (1955) for the mass range
0.4$<$M/M$_{\odot}$$\le$10.

\begin{figure}[h]
\begin{center}
\includegraphics[height=0.40\textheight,width=0.8\textwidth]{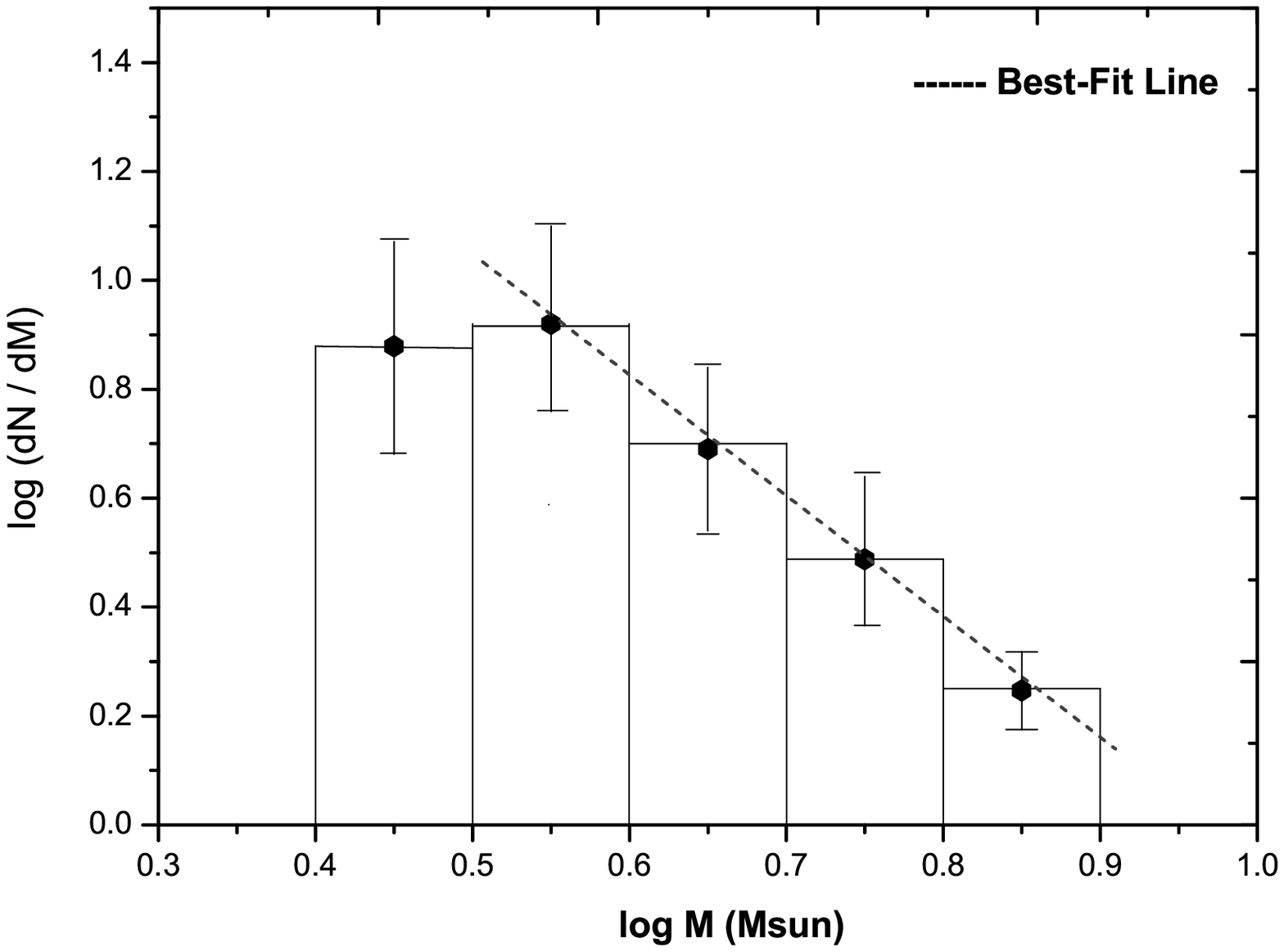}
\caption{ Mass distribution function for YSOs in Sh2-82. The error
bars represent $\pm$ $\sqrt{N}$ errors.}
\end{center}
\end{figure}

\section{Discussion}
 Although Sh2-82 is seen at the edge of
the pulsar wind nebula (PWN) G54.1+0.3, it is not related to the
PWN, the center of which has a distance of at least 5kpc (Weisberg
et al. 2008). G54.1+0.3 is likely to be at a distance of
$\approx6.2$ kpc due to the morphological association of the PWN
with a CO molecular cloud at a velocity of $\approx53$ km s$^{-1}$,
as revealed by high-resolution $^{13}$CO images(Leahy et al. 2008).
We now discuss the likelihood of triggering star formation in
Sh2-82. According to the model of Elmegreen \& Lada (1977), a thin
layer of compressed neutral material forms between a shock front and
a slow-moving D-type ionization front. This may be the configuration
observed in Sh2-82. In the triggering star-formation scenario, the
ages of the ionizing star(s) must be older than the ages of the
second generation stars plus the shock front traveling time.
However, a problem with Elmegreen \& Lada's model concerns the ages
and sizes of the HII regions.

Using the model described by Dyson $\&$ Williams (1980), we
calculate the dynamical age of the HII region at a given radius R as
\begin{equation}
t(R)=\frac{4 R_{s}}{7 c_{s}} [(\frac{R}{R_{s}})^{7/4}-1]
\end{equation}
where c$_{s}$ is the sound velocity in the ionized gas (c$_{s}$=10
km s$^{-1}$) and R$_{s}$ is the radius of the Str\"{o}mgren sphere
given by R$_{s}$= (3N$_{uv}$/4$\pi$n$^2$$_0$$\alpha$$_B$ )$^{1/3}$,
where N$_{uv}$ represents the Lyman continuum photons emitted by the
ionizing star per second, and $\alpha$$_B$ = 2.6 $\times$ 10$^{-13}$
cm$^3$ s$^{-1}$ is the hydrogen recombination coefficient. Sh2-82 is
excited by a B0V star, which emits 1.25$\times$10$^{48}$ ionizing
photons per second (Vacca et al. 1996; Schaerer $\&$ de Koter 1997).
The present radius of Sh2-82 could be calculated to be 2.2$\pm
$0.5pc at a distance of 1.7$\pm$ 0.4kpc. We derived a dynamical age
of between 0.25 and 0.48 Myr, assuming an original ambient density
of (1 $\pm$ 0.5) $\times$ 10$^{3}$cm$^{-3}$.

Whitworth et al. (1994,; hereafter W94) presented an analytical
treatment of the collect-and-collapse process. A shock front forms
and gathers material until it is able to fragment and collapse to
form stars. The fragmentation time and radius can be calculated as:

\begin{equation}
t_{frag}=1.56 Myr\times a_{0.2}^{7/11} L_{49}^{-1/11} n_{3}^{-5/11}
\end{equation}
\begin{equation}
R_{frag}=5.8 pc\times a_{0.2}^{4/11} L_{49}^{1/11} n_{3}^{-6/11}
\end{equation}
where a$_{0.2}$ is the sound speed inside the shocked layer in units
of 0.2 km s$^{-1}$, L$_{49}$ is the central source ionizing flux in
units of 10$^{49}$photons s$^{-1}$, and n$_{3}$ is the initial gas
number density in units of 10$^{3}$cm$^{-3}$. Since we cannot
calculate an appropriate value for a$_{0.2}$ from our current data
sets, we adopt a value of 1.0 as Koenig et al.(2008) did in the
analysis of W5. We find that the fragmentation process in the
periphery of Sh2-82 should occur between 1.6 and 5.3 Myr after its
formation, a later point in time than its dynamical age derived
above. We thus discard the so-called collect and collapse as being
the mechanism responsible for YSO formation. Other processes such as
radiative-driven implosions and/or collisions with pre-existing
molecular core(s) may operate; we can see from figure 1 that the CO
emission overlaps the densest part of the MSXA shell, which may be
evidence of the radiation driven. Sh2-82 is not the only object; HII
regions like Sh2-235 (Kirsanova et al. 2008), Sh2-217 and Sh2-219
(Deharveng et al. 2003) also show such physical processes of
sequential star formation.

\section{Summary}
The interstellar medium and its young stellar objects around the HII
region Sh2-82 have been analyzed. Two molecular clumps were found
from the archival data of GRS. We found two young corresponding
clusters with 2MASS embedded in the two molecular clumps. By using
spectral energy distribution fitting to each of the YSO candidates
with infrared access, we derived the mass function slopes in this
region. Comparing the HII region dynamical age and the fragmentation
time of the molecular shell, we discarded the so-called collect and
collapse as being the triggering process for YSO formation. Sh2-82
can be a mixture of other processes such as radiative-driven
implosion and/or collisions with pre-existing clumps. More numerical
studies and deeper observations should be carried out in the future
to reveal an even clearner picture of triggered star formation in
HII regions.

\section*{Acknowledgement}
This publication makes use of molecular line data from the Boston
University-FCRAO Galactic Ring Survey (GRS). The GRS is a joint
project between Boston University and the Five College Radio
Astronomy Observatory, funded by the National Science Foundation
under grants AST-9800334, AST-0098562, AST-0100793, AST-0228993, \&
AST-0507657. We are also grateful to the anonymous referee for whose
constructive suggestions.

   \end{document}